\documentclass[preprint,amsmath,amssymb,superscriptaddress]{revtex4}%
\usepackage{graphicx}
\usepackage{dcolumn}
\usepackage{bm}
\usepackage{amsmath}
\usepackage{amsfonts}
\usepackage{amssymb}
\usepackage{epstopdf}
\usepackage{color}%
\begin{document}

\title{Landau-Zener-St\"{u}ckelberg-Majorana interference in a 3D transmon
driven by a {chirped} microwave}
\author{Ming Gong}
\thanks{These authors contributed equally to this work.}
\affiliation{National Laboratory of Solid State Microstructures, School of Physics,
Nanjing University, Nanjing 210093, China}
\affiliation{Department of Physics and Astronomy, University of Kansas, Lawrence, KS
66045, USA}
\author{Yu Zhou}
\thanks{These authors contributed equally to this work.}
\affiliation{Research Institute of Superconductor Electronics, School of Electronic\\
Science and Engineering, Nanjing University, Nanjing 210093, China,}
\author{Dong Lan}
\affiliation{National Laboratory of Solid State Microstructures, School of Physics,
Nanjing University, Nanjing 210093, China}
\author{Yunyi Fan}
\affiliation{Research Institute of Superconductor Electronics, School of Electronic\\
Science and Engineering, Nanjing University, Nanjing 210093, China,}
\author{Jiazheng Pan}
\affiliation{Research Institute of Superconductor Electronics, School of Electronic\\
Science and Engineering, Nanjing University, Nanjing 210093, China,}
\author{Haifeng Yu}
\email{hfyu@nju.edu.cn}
\affiliation{National Laboratory of Solid State Microstructures, School of Physics,
Nanjing University, Nanjing 210093, China}
\affiliation{Synergetic Innovation Center of Quantum Information and Quantum Physics,
University of Science and Technology of China, Hefei, Anhui 230026, China,}
\author{Jian Chen}
\affiliation{Research Institute of Superconductor Electronics, School of Electronic\\
Science and Engineering, Nanjing University, Nanjing 210093, China,}
\author{Guozhu Sun}
\email{gzsun@nju.edu.cn}
\affiliation{Research Institute of Superconductor Electronics, School of Electronic\\
Science and Engineering, Nanjing University, Nanjing 210093, China,}
\affiliation{Synergetic Innovation Center of Quantum Information and Quantum Physics,
University of Science and Technology of China, Hefei, Anhui 230026, China,}
\author{Yang Yu}
\affiliation{National Laboratory of Solid State Microstructures, School of Physics,
Nanjing University, Nanjing 210093, China}
\affiliation{Synergetic Innovation Center of Quantum Information and Quantum Physics,
University of Science and Technology of China, Hefei, Anhui 230026, China,}
\author{Siyuan Han}
\affiliation{Department of Physics and Astronomy, University of Kansas, Lawrence, KS
66045, USA}
\author{Peiheng Wu}
\affiliation{Research Institute of Superconductor Electronics, School of Electronic\\
Science and Engineering, Nanjing University, Nanjing 210093, China,}
\affiliation{Synergetic Innovation Center of Quantum Information and Quantum Physics,
University of Science and Technology of China, Hefei, Anhui 230026, China,}

\begin{abstract}
By driving a 3D transmon with microwave fields, we generate an effective
avoided energy-level crossing. Then we chirp microwave frequency, which is
equivalent to driving the system through the avoided energy-level crossing by
sweeping the avoided crossing. {A double-passage chirp produces 
Landau-Zener-St\"{u}ckelberg-Majorana interference that
agree well with the numerical results.} Our method is fully applicable to
other quantum systems that contain no intrinsic avoided level crossing,
providing an alternative approach for quantum control and quantum simulation.
\end{abstract}

\maketitle

Landau-Zener-St\"{u}ckelberg-Majorana (LZSM) interference \cite%
{Landau1932a,Landau1932b,Zener1932,Stuckelberg1932,Francesco2005} is a well
known quantum phenomena, resulting from sweeping a system back and forth
across an avoided energy-level crossing in the energy diagram. It has been
extensively explored in a lot of systems \cite{Nori2010} including atomic
systems \cite{Van2009}, quantum dots \cite{Ribeiro2009}, and superconducting
qubits \cite{Oliver2005,Silla2006,Izmalkov2008,Sun2009,SunNat.Commun.}. In the recent
practice of quantum information processing, LZSM interference provides a
useful tool to calibrate some crucial characteristics of a system, e.g., the
energy-level structure, the coupling strength of the quantum states, and the
decoherence time \cite{Shytov2003,Oliver2005,Silla2006}. It also finds applications in manipulating quantum states.
Conventionally, in order to generate LZSM interference one has to locate an
avoided energy-level crossing in the energy diagram of the system. Then one
sweeps the external parameter to drive the system across the avoided
energy-level crossing, where Landau-Zener (LZ) transition occurs. The
split states of the system evolve along two different paths,
accumulating phase difference. When one sweeps the system back and pass the
avoided energy-level crossing again, the split states will interfere,
creating LZSM interference patterns. However, for some quantum systems,
there is no avoided energy-level crossing in the energy diagram. Even worse,
their energy-level spacings may be independent of the
external bias parameters therefore one cannot drive the system by sweeping
the external parameters. A typical case with these two properties is a 3D
transmon \cite{Paik2011,Rigetti2012}, which is an improved version of a
superconducting qubit \cite{Clarke2008}. Although significant
amounts of quantum phenomenon have been demonstrated \cite{Devoret2013,campagne2013persistent,kirchmair2013observation,murch2013observing}, LZSM interference
has not been reported in 3D transmon so far.

{Chirping field is a widely used method for coherent population transfer in atomic and molecular systems 
\cite{broers1992efficient, chelkowski1990efficient, melinger1994adiabatic, cao1998molecular}. }
In this letter, we realize LZSM interference in a 3D transmon by using
microwave with {chirped} frequency. In the rotating frame, a microwave driven
3D transmon exists an effective avoided energy-level crossing {\cite{Sun2011}}%
. Gradually chirping microwave frequency, we can drive the system through
the avoided energy-level crossing instead of sweeping the external bias parameter \cite{zhou2014observation}. By sweeping the system twice across the avoided energy-level crossing, we observe LZSM interference and show the quantum dynamic evolution of LZSM interference changing with the initial states and detuning. The
numerical simulated results agree with the experimental data very well.

The sample we used is a transmon qubit centered in a 3D rectangular aluminum
(Al 6061-T6 alloy) cavity. The fundamental resonant frequency of the bare cavity is 9.0131 GHz. The transmon is fabricated with
standard double-angle shadow evaporation of aluminum on a high-resistivity
silicon substrate. The sample is mounted on the mixing chamber of a dilution
refrigerator with base temperature about 20 mK. A $\mu $-metal can is used
to shield the external magnetic field. In order to achieve high SNR, the
input signal is heavily attenuated and the low-noise microwave amplifiers
have been used for the output signal \cite{Rigetti2012}. The qubit state is
readout by a standard \textquotedblleft bright state\textquotedblright\
readout technique of heterodyne technique \cite{Reed2010}.

\begin{figure}[ptb]
	\includegraphics[width=0.5\textwidth]{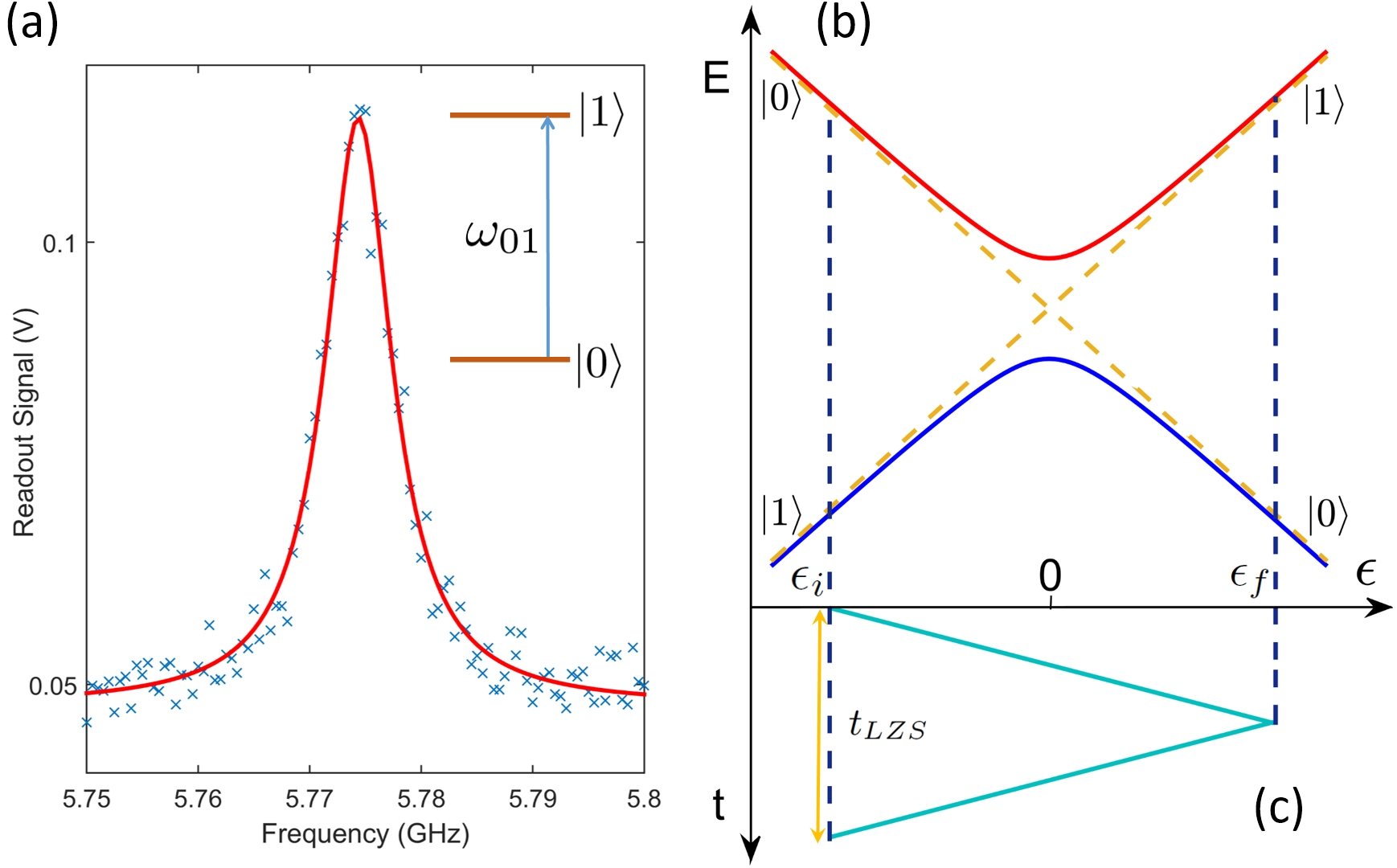}
	\caption{\textbf{(a)} The spectroscopy of the qubit, with the resonant qubit frequency $\omega_{01}/2\pi$=5.7744 GHz. The inset
		is a diagram of the energy levels of a 3D transmon qubit. \textbf{(b)}
		Schematic diagram of the energy-level of a 3D transmon qubit driven by a {chirped} microwave. \textbf{(c)} Time profile of $\epsilon$ in performing LZSM interference.}
	\label{fig:epsart}
\end{figure}

The truncated Hamiltonian of the two lowest levels $|0\rangle $ and $%
|1\rangle $ of a 3D transmon qubit are $H_{0}=\hbar \omega _{01},$ where $\omega _{01}$ is
the energy difference between $|0\rangle $ and $|1\rangle $. From spectroscopy measurement we obtain that $\omega _{01}/2\pi$ = 5.7744 GHz, as shown in Fig. 1(a).
{With the second excited energy level determined as $\omega _{02}/2\pi$ = 11.2744 GHz , we calculated the Josephson coupling energy as $E_{J}/h= 16.7\pm 0.1$ GHz, and the charge
energy as $E_{C}/h= 274\pm2$ MHz \cite{Koch2007}}.

If we drive the qubit with microwave $A_{r}$sin$\omega _{r}t,$ the
Hamiltonian of the driven system is identical to a quantum two-level system (%
$\hbar \equiv 1$)
\begin{equation}
H=-\frac{1}{2}(\epsilon \sigma _{z}+\Omega _{0}\sigma _{x}),
\end{equation}%
where $\epsilon =\omega _{r}-\omega _{01}$ is the detuning and $\Omega _{0}$
is the gap size of the avoided energy-level crossing, as shown in Fig. 1(b),
which is proportional to the amplitude of the microwave field $A_{r}$. In
general, for a 3D transmon, $\omega _{01}$ is constant. We cannot change $%
\omega _{01}$ to generate ordinary LZ transition by sweeping the external field. However, in order to sweep the system across the avoided
energy-level crossing, we can chirp the microwave frequency $\omega _{r}$
from $\omega _{i}$ to $\omega _{f}$ \ linearly to cross $\omega _{01}$,
i.e., $\epsilon (t)=\omega _{r}(t)-\omega _{01}$. In our experiments, the
chirp operation is easily realized by applying intermediate frequency (IF)
modulation signals generated by an arbitrary waveform generator (Tektronix
70002) to the I/Q ports of a vector signal generator (R$\&$S SGS100A).
{Suppose the local microwave signal is $A_{r}\sin (\omega _{01}t)$, the waveforms
applied on the I and Q ports are quadrature signals, i.e., $\cos (\delta
_{\omega }t+\phi_0)$ and $\sin (\delta _{\omega }t+\phi_0)$, respectively, then the modulated
microwave waveform is $A_{r}\sin
((\omega _{01}+\delta _{\omega })t+\phi_0)$, where $\delta _{\omega }=\epsilon =vt$, $v$ the sweeping speed, and $\phi_0$ the initial phase of the modulation.
In our experiment, we set $\phi_0=0$ so that the coupling between the {chirped} microwave 
field and the qubit is in $x$ direction.}
A {chirp} operation with same speed is performed to sweep the system back across
the avoided level crossing, as shown in Fig. 1(c). The double passage
passing the avoided crossing leads to LZSM interference, which is analogous
to Mach-Zehnder interferometry in optics \cite{Oliver2005}. 
%It is worth to emphasize that the initial phase of the second modulation signals needs to be 
%the same as the final phase of the first modulation signals to ensure that the coupling of
%the microwave and the qubit is kept at the $x$ direction. 
After {the chirp
operation}, a state tomography measurement \cite{Steffen2006,Leek2007,deng2015observation} is
performed to obtain the expectation value of $\langle \sigma _{x,y,z}\rangle
$ of the qubit state.

We investigate LZSM interference for various initial states and detuning.
First of all, we choose the gap size of the avoided crossing as $%
\Omega _{0}/2\pi =20$ MHz and initialize the state in $|0\rangle $ at $\epsilon _{i}/2\pi=-400$
MHz, satisfying $|\epsilon _{i}|/\Omega _{0}=20\gg 1,$ which indicates
that the sweeping starts far away from the center of avoided crossing. Then
we can adjust the sweeping range $\epsilon _{f}$ and sweeping time $t_{LZ}$,
which are defined by the final frequency of chirp and the chirped speed,
respectively. In order to produce LZSM interference, we let the
qubit pass the avoided crossing twice. For simplicity the sweeping
speed keeps unchanged for the two passages. Therefore, the whole time for LZSM
interference evolution is $t_{LZS}=2t_{LZ}.$ Shown in Fig. 2(a) are the
typical patterns of LZSM interference, where $%
t_{LZ}$ varies from $1$ ns to $50$ ns, and $\epsilon
_{f}/2\pi $ varies from $-400$ MHz to $400$ MHz. All three components of the
qubit state represented by the expectation values of $\langle \sigma
_{x,y,z}\rangle $ are measured by performing the state tomography
measurement after the LZSM evolution.

In order to confirm the observation, we compare the results with those of
the numerical simulation. The quantum dynamics of system can be described
with the master equation of the time evolution of the density matrix $\rho $
considering the effects of dissipation
\begin{equation}
\dot{\rho}=\frac{1}{i\hbar }[H,\rho ]-\Gamma \lbrack \rho ],
\end{equation}%
where $H$ is the Hamiltonian of the system given by Eq. (1). $\Gamma \lbrack
\rho ]$ describes the decoherence effect in the evolution
phenomenologically, including the relaxation time $T_{1}$ and dephasing time
$T_{2}^{\ast }$. {By substituting $T_{1}= 2.38 \pm 0.13$
$\mu s$ determined from energy relaxation measurement and $T_{2}^{\ast }= 2.27 \pm 0.33$ $\mu s$ determined from Ramsey fringe measurement}, we obtain the numerical
patterns, as shown in the insets of Fig. 2(a). The agreement between
the theoretical and experimental results are excellent, {indicating the validity
of our chirp method in realization of LZSM interference.}

\begin{figure}[ptb]
	\includegraphics[width=0.5\textwidth]{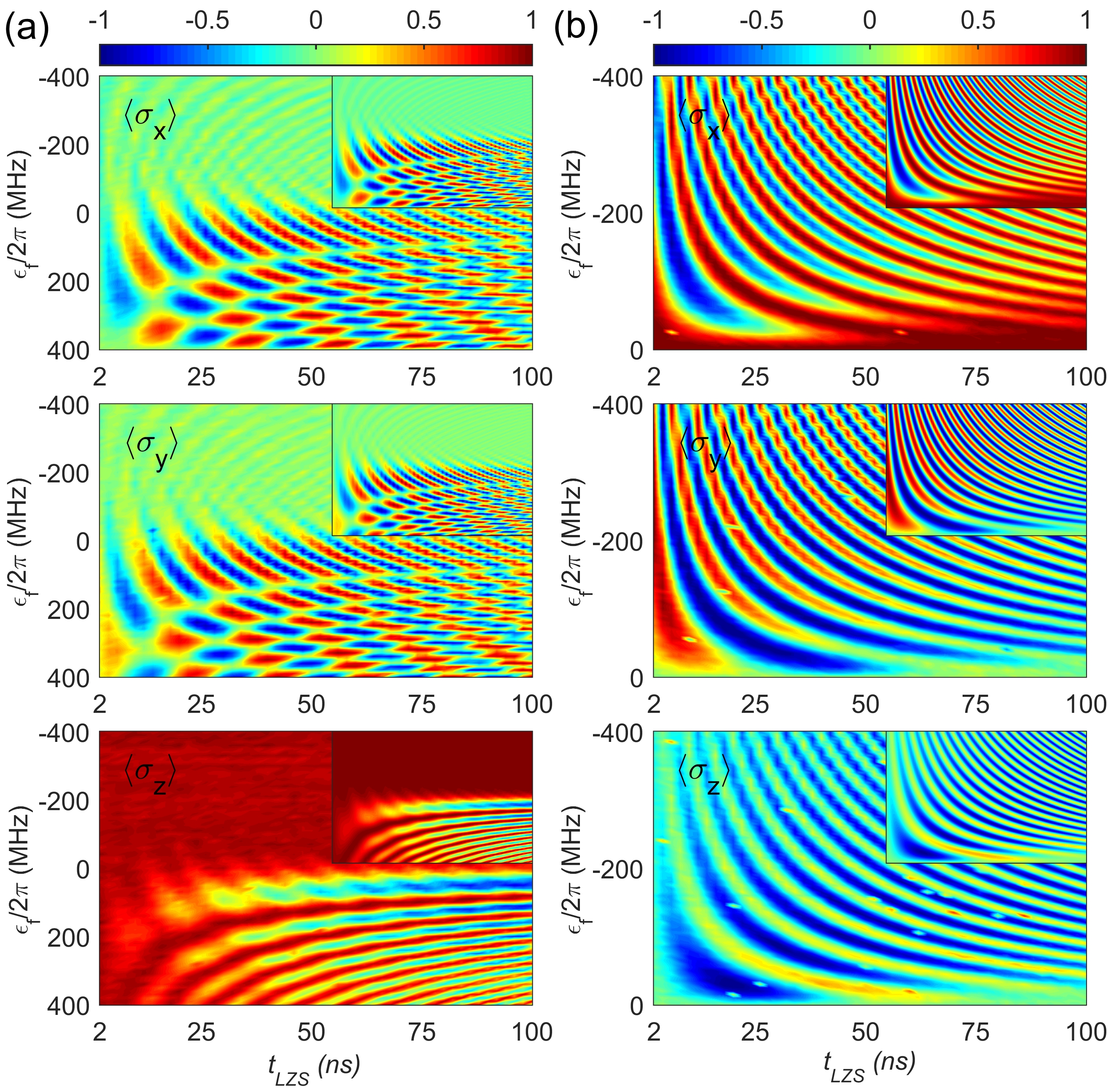}
	\caption{\textbf{(a)} and \textbf{(b)} Measured expectation values $\langle \protect\sigma_{x,y,z} \rangle$ of the
		qubit state as a function of $\protect\epsilon_f/2%
		\protect\pi$ and $t_{LZS}$. The insets are the numerical
		simulations. In (a), the initial state is far away from the avoided energy-level crossing, prepared in $|0\rangle$ at $%
		\protect\epsilon_i/2\protect\pi=-400$ MHz. In (b), the initial state is in the center of
		the avoided energy-level crossing, prepared in
		$\frac{1}{\protect\sqrt{2}}(|0\rangle+|1\rangle)$ at $\protect\epsilon_i/2%
		\protect\pi=0$. Almost identical LZSM interference patterns are observed in the experimental and numerical results.}
	\label{fig:epsart}
\end{figure}

For another case, we prepare the initial state in $\frac{1}{\sqrt{2}}(|0\rangle +|1\rangle )$, an energy eigenstate of
the system, by applying a resonant $-\pi /2$
rotation around the $y$ axis. This time we start the evolution from the
center of the avoided crossing, i.e., $\epsilon _{i}/2\pi =0$. Similar to the
previous experiment, we choose $\Omega _{0}/2\pi =20$ MHz, $\epsilon _{f}/2\pi $ ranges from $0$ to $-400$ MHz, and $t_{LZ}$ varies from $1$ ns to $50$ ns. The results of the
expectation values $\langle \sigma _{x,y,z}\rangle $ of the qubit state are
shown in Fig. 2(b), which are also in excellent agreement with the numerical
simulations (insets). We would like to mention that LZSM interferences in
previous work usually start sweeping parameter far away from the avoided
crossing. Although it is not difficult to theoretically calculate the
evolution starting from the center of the avoided crossing with the initial
state being the energy eigenstate, there are two obstacles for the
experimental investigation. One is how to define the exact center of the
avoided crossing. The other is how to create the high fidelity initial
eigenstate which is the superposition of diabatic states. Our {chirp}
method solves both problems simultaneously: The avoided crossing is easily
defined at $\omega _{r}(t)=\omega _{01}.$ At the same time, the initial state can be prepared
with on-resonant microwave therefore one can initialize the state to any
point on Bloch sphere, including the energy eigenstate of the system.

\begin{figure}[ptb]
	\includegraphics[width=0.5\textwidth]{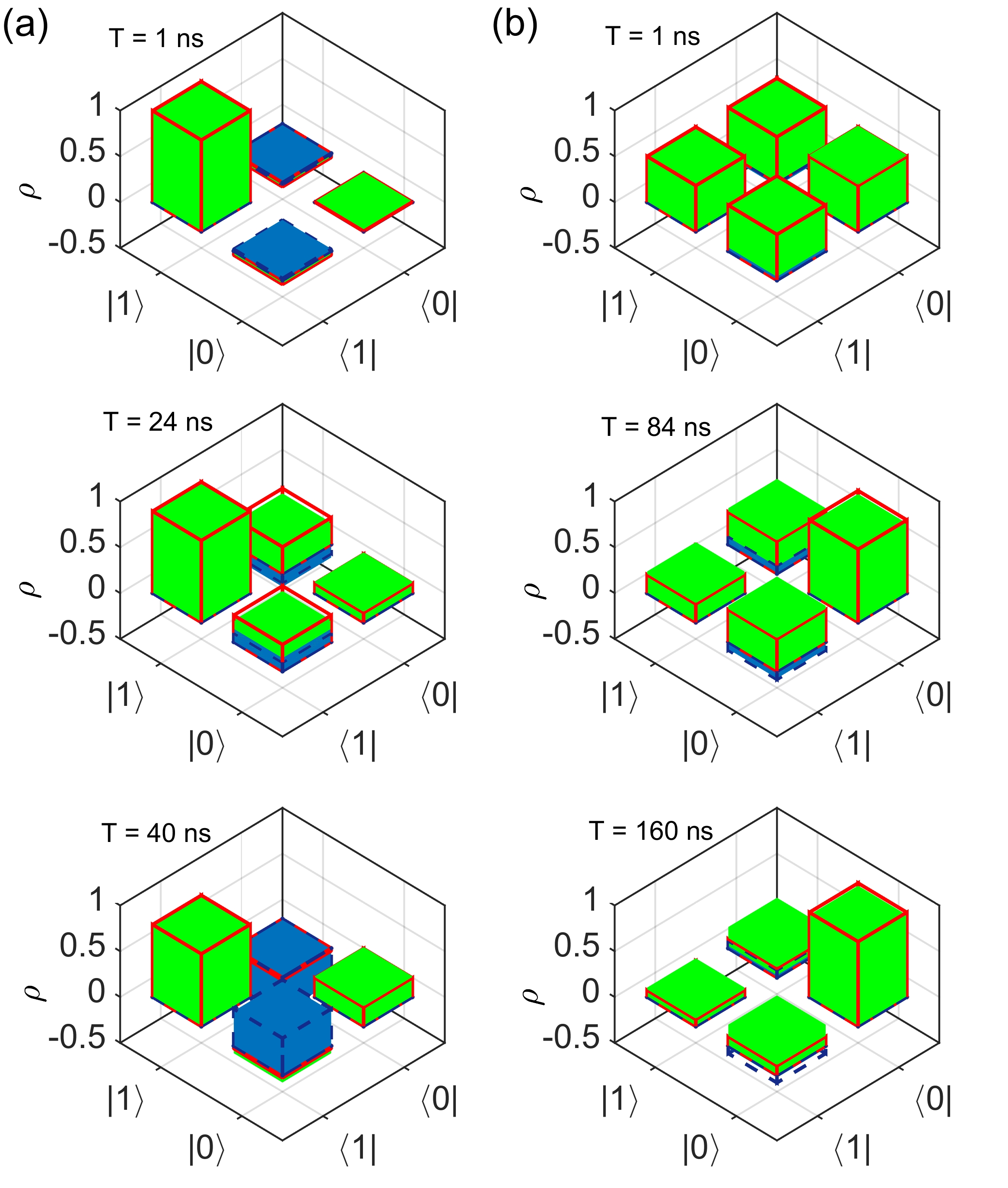}
	\caption{\textbf{(a)} and \textbf{(b)} State tomography of the qubit in the evolution of LZSM interference. The real (image) parts of the
		experimental and numerical simulated density matrix are shown as green (blue) solid bars and red solid (blue dashed) lines, respectively.
		In (a), the initial state is prepared in $|0\rangle$ at $%
		\protect\epsilon_i/2\protect\pi=-400$ MHz. The total evolution time is $40$ ns and $\protect\epsilon_f/2\protect\pi$ is
		$400 $ MHz. The density matrix of the qubit at T $=1$ ns, $24$ ns and $40$ ns
		are presented, respectively.	
		In (b), the initial state is prepared in
		$\frac{1}{\protect\sqrt{2}}(|0\rangle+|1\rangle)$ at $\protect\epsilon_i/2%
		\protect\pi=0$. The total evolution time is $160$ ns and $\protect\epsilon_f/2\protect\pi$ is
		$-200 $ MHz. The density matrix of the qubit at T $=1$ ns, $84$ ns and $160$ ns
		are presented, respectively.
	}
	\label{fig:epsart}
\end{figure}

The excellent agreement between the experimental data and the simulation
results indicates that with a {chirp} technique one can realize and
investigate LZSM interference completely in a system without intrinsic
avoided energy-level crossings. The splitting of the avoided crossing and
the sweeping range can be tuned conveniently by changing the {chirp}
parameters such as $\Omega _{0}$ and $\delta _{\omega }$. We are able to prepare the initial state
completely in one of the eigenstates without any leakage to the other one.

Using the method of {chirping} frequency, one can also perform time-resolved
state tomography measurement to obtain the dynamical evolution of the qubit during LZSM
interference. We set $\epsilon _{i}/2\pi =-400$ MHz
and $\epsilon _{f}/2\pi =400$ MHz. The total evolution time is $40$ ns. At
each step which is $1$ ns, the state of the qubit is measured. Then the
density matrix in the dynamical evolution of the qubit in LZSM interference
can be obtained. For example, the density matrix of the qubit at evolution time T $=1$ ns, $24$ ns and $40$ ns
are shown in Fig. 3(a), representing the qubit state at the beginning, after
the first LZ transition, and at the end of a LZSM interference process, respectively.
The state evolution confirms the physics picture of LZSM interference, in which the system
starts from the diabatic state, separates to a superposition state after the first
LZ transition, then interferes at the second LZ transition.
Similarly, we also investigate the evolution starting from the center of avoided
crossing, i.e., $\epsilon _{i}/2\pi =0$. Here $\epsilon _{f}/2\pi =-200$ MHz and $t_{LZS}=160$ ns. In this situation, the system starts from the superposition state and interferes when it passes the avoided crossing. We perform the state tomography measurement in the process
to obtain the density matrix of the qubit at T $=1$ ns, $84$ ns and $160$ ns shown in Fig. 3(b), representing the qubit state at the beginning, middle, and end of the evolution. The evolution of the density matrix clearly show the process mentioned above.
Numerical simulations for both cases agree well with the measured evolution of density
matrix.

Compared with the conventional method of realizing LZSM interference by
sweeping external bias parameter, our {chirp} method has several
advantages. The first one is that during the evolution, all parameters of
the qubit, such as $T_{1}$, $T_{2}^{\ast }$ and the coupling strength
between the qubit and the external driving field, keep unchanged in the
whole process. These parameters usually depend on the external bias.
Therefore, when we generate LZSM interference by sweeping external bias, the
evolution may be complicated. Secondly, the {chirped} range $\epsilon _{i}$
and $\epsilon _{f}$ in our method is not limited by the structure of the
qubit energy diagram. For instance, it is not affected by the nearby intrinsic avoided energy-level crossings or the additional splittings caused
by the coupling to the microscopic two-level systems. Thirdly, it is easy to
control the {chirped} velocity and the coupling strength between the driving
field and qubit by controlling the microwave frequency and power.

In summary, we realize LZSM interference in a superconducting 3D transmon
which has constant energy level spacing thus containing no intrinsic avoided
energy-level crossing.
An effective avoided crossing is created by the microwave field. Then
we chirp the microwave frequency and drive the system through the avoided
crossing. By sweeping the system twice across the avoided crossing, we
observe LZSM interference. Our method can
be applied to the systems whose energy diagrams lack intrinsic
avoided energy-level crossings and/or can not be changed rapidly by sweeping
external parameters. As long as they can interact with external microwave
irradiation, one can generate LZSM interference to calibrate some crucial
characteristics of the system and to conduct quantum control and/or quantum
simulation.

This work is partially supported by the SKPBR of China (2011CB922104), NSFC
(11474154, 93121310, 11274156, 61521001, BK2012013, 11474152), PAPD, a doctoral
program (20120091110030) and Dengfeng Project B of Nanjing University.

%\bibliographystyle{unsrt}
%\bibliographystyle{plain}
%\bibliography{LZSM_Ref}
%{PDSR}

\newpage

\end{document}